\documentstyle[12pt]{article}

\begin{document}
\begin{titlepage}
\begin{center}

\today     \hfill    LBNL-53835 \\

\vskip .5in

{\large \bf Correspondence and Analyticity.}
\footnote{This work is supported in part by the Director, Office of Science, 
Office of High Energy and Nuclear Physics, Division of High Energy Physics, 
of the U.S. Department of Energy under Contract DE-AC03-76SF00098}

\vskip .50in
Henry P. Stapp\\
{\em Lawrence Berkeley National Laboratory\\
      University of California\\
    Berkeley, California 94720}
\end{center}

\vskip .5in

\begin{abstract}
The analyticity properties of the S-matrix in the physical region are 
determined by the correspondence principle, which asserts that the 
predictions of classical physics are generated by taking the classical limit 
of the predictions of quantum theory. The analyticity properties deducible 
in this way from classical properties include the locations 
of the singularity surfaces, the rules for analytic continuation around 
these singularity surfaces, and the analytic character (e.g., pole, 
logarithmic, etc.) of these singularities. These important properties 
of the S-matrix are thus derived without using stringent locality assumptions,
or the Schroedinger equation for temporal evolution, except for freely 
moving particles. Sum-over-all-paths methods that emphasize paths of 
stationary action tend to produce the quantum analogs of the contributions 
from classical paths. These quantum analogs are derived directly from the 
associated classical properties by reverse engineering the 
correspondence-principle connection.
     
(This article is an invited contribution to a special issue of Publications
of RIMS commemorating the fortieth anniversary of the founding of the 
Reseach Institute for Mathematical Science.)

\end{abstract}

\end{titlepage}

\newpage
\renewcommand{\thepage}{\arabic{page}}
\setcounter{page}{1}

{\bf 1. Introduction.}

The S matrix was introduced by Wheeler[1]. It specifies the amplitude for 
the scattering of any set of originally noninteracting initial particles 
to any set of eventually noninteracting final particles. The full physical 
content of relativistic quantum theory resides in the S matrix: any two 
formulations that give the same S matrix are considered to be physically 
equivalent.

The S matrix is a function of the momentum-energy four-vectors
of the initial and final particles. The law of conservation of momentum-energy
entails that the term of the S matrix that describes the scattering of any 
specified set of initial particles to any specified set of final particles 
must have a momentum-energy conservation-law delta function that constrains 
the sum of the momentum-energy vectors of the final particles to be 
equal to the sum of the momentum-energy vectors of the initial 
particles. The remaining factor, which is defined only at points that satisfy 
this conservation-law condition, is called a scattering function. It is finite
at almost all points in its domain of definition. This is important because 
computations starting from the Schroedinger equation tend to give scattering
functions that are everywhere infinite. Thus Heisenberg[2] and others [3] have 
proposed an approach to relativistic quantum theory that avoids the infinities
that arise from the Schroedinger equation by discarding that equation 
altogether, and computing the S matrix directly from certain of its general 
properties. In this approach one never specifies the (Schroedinger-equation-
induced) temporal evolution that takes initial states continuously to final 
states, but which, according to the basic philosophy, lacks physical 
significance. The S-matrix method works very well for simple cases. It may 
work in general, but new computational techniques would be needed to achieve 
this.

A key property of the scattering functions is that each of them is 
analytic (holomorphic) at almost every point of its original (real) domain 
of definition. This property was originally deduced from an examination of 
Feynman's formulas for these functions, which are derived essentially from 
the (relativistic) Schroedinger equation. Landau[4] and Nakanishi[5] 
independently deduced the very restrictive necessary conditions for the 
occurrence of singularities of these functions. Coleman and Norton[6] then 
noted that these Landau-Nakanishi conditions are precisely the conditions 
for the existence of a {\it classical} physical process that has the same 
topological structure --- i.e., has the same arrangement of line segments 
connected at vertices --- as the Feynman graph with which it is associated.  

A Feynman graph is topological structure of line segments joined at vertices.
It was used by Feynman to specify a corresponding mathematical contribution 
to the S matrix. The associated Landau-Nakanishi diagram is a diagram in 
four-dimensional space-time that has the same topological structure, but 
moreover satisfies all of the conditions of a corresponding process in 
classical physics. Thus a Landua-Nakanishi diagram can be regarded as a 
representation of a process in classical-physics that consists of a network 
of point particles that interact only at point vertices, and that propagate 
between these vertices as freely moving particles.

The rules of (relativistic) classical particle physics assign a momentum-energy
four-vector to each line of the diagram, and impose the conservation-law
condition that the energy-momentum flowing into the diagram along the 
initial incoming lines must be able to flow along the lines of the graph, 
and then out along the final outgoing lines {\it with energy-momentum 
conserved at each vertex.} This conservation-law condition is imposed also 
by the Feynman rules. But the Landau-Nakanishi (i.e., classical-physics) 
diagram is required to satisfy also the ``classical physics'' requirement 
that each line of the spacetime diagram be a {\it straight-line segment that 
is  parallel to the momentum-energy carried that line.} [In classical 
relativistic particle physics each freely-moving particle moves in space-time 
in the direction of its momentum-energy four-vector ($p=mv,v^2=1$), but this 
property is not imposed in quantum theory: it would conflict with the 
uncertainty principle, and, likewise,  with the Fourier-transformation 
connection between space-time displacements and momentum-energy that 
constitutes the foundation of quantum theory.] 

The Landau-Nakanishi diagram is, then, the picture of a possible classical
process, involving point particles interacting at points, and conforming to 
the conditions of relativistic classical-particle physics. These conditions 
were shown by Landau and Nakanishi to specify the {\it location (in the space
of the momentum-energy four vectors of the initial and final particles) of 
a singularity---failure of analyticity---of the contribution to the S matrix 
corresponding to the associated Feynman graph.}

The purpose of this article is to highlight the fact that although this 
important connection between the physical-region singularities of the quantum 
scattering functions and associated classical scattering processes was 
originally derived from very strong quantum assumptions involving the 
concepts of point interactions and continuous Schroedinger evolution in 
time, the result is actually a consequence of much less. It is a consequence 
of the ``correspondence principle'' connection between relativistic quantum 
physics and relativistic classical-particle physics. This principle asserts 
that the predictions of classical physics emerge from quantum theory 
in the ``classical limit'' in which all effects due to the nonzero value 
of Planck's constant become negligible. 

The correspondence principle entails, however, much more than just the 
analyticity of the S matrix at all points that do not correspond to a 
classical-physics process. It entails also that, in a real neighborhood 
of almost every real singular point, the scattering function is the limit 
of a function analytic in the interior of a certain cone-like domain that 
extends some finite distance into the complex domain from its tip in the 
real neighborhood. This means that {\it each physical scattering function 
is a limit of single analytic function.} That feature of the S matrix is 
one of the key general properties upon which the S-matrix approach is based. 
Its derivation from the correspondence principle was given by Chandler and 
Stapp[7] and by Iagolnitzer and Stapp[8]. The first of these two papers sets 
out the general framework, but is formulated within a distribution-analytic 
framework in which the wave functions are, apart from mass-shell-constraint 
delta functions, infinitely differentiable functions of compact support. 
Consequently, it achieves analyticity only modulo infinitely differentiable 
background terms. The second of these papers uses essentially Gaussian 
wave functions to obtain full analyticity.      

It is worth noting that Sato [9] independently constructed a mathematical
machinery called the sheaf of microfunctions, which can be used to describe
the same cone-like domain when applied to the S matrix.

The correspondence principle entails even more. It specifies also the 
{\it nature} of these singularities: whether they are, for example, pole,  
or logarithmic singularities. This means that the quantum effects
closely associated with these classical-physics processes are determined 
already by the correspondence principle, without appeal the notion of
true point interactions or of the Schroedinger equation. That is, the 
correspondence principle, which is a condition on the classical limit of 
quantum theory, can be ``reverse engineered'' to deduce those features of the 
quantum S matrix that produce the classical result in the classical limit.
And these feature include the analytic character of the the S Matrix 
scattering functions in their original (real) domains of definition.

{\bf 2. An Asymptotic Fall-Off Property.}

The papers with Chandler and Iagolnitzer just cited deal exclusively with 
particles of non-zero rest mass. The momentum-space wave function of particle 
$i$ then has, due to the mass-shell condition, the form
$$
\Psi_i (p_i) = \psi_i (p_i) 2\pi \delta (p_i^2 -m_i^2),   \eqno (2.1)  
$$
where $p_i^2$ is the Lorentz inner product of $p_i$ with itself, with metric
$(1,-1,-1,-1)$, and $m_i$ is the (nonzero) rest-mass of particle $i$.  
Quantum theory is characterized, fundamentally, by the Fourier-transform 
link between momentum-energy and space-time. Thus the spacetime form of 
this momentum-energy wave function is given by the Fourier transform:
$$ 
\widetilde\Psi_i (x_i) = \int (2\pi)^{-4} d^4 p_i \exp (-ip_ix_i)\Psi_i (p_i). 
\eqno (2.2)
$$

The spacetime wave function has important asymptotic fall-off properties. 
In Appendix A of reference [13] it is shown that if $\psi_i(p_i)$ has compact
support and is continuous, together with its first and second derivatives,
and if $u$ is any positive time-like four-vector satisfying $v^2 =1$, then
$$
\lim_{\tau \rightarrow \infty} f(m_i,\tau)\widetilde\Psi_i(v\tau)
      =\psi_i(m_iv),   \eqno (2.3)  
$$
where 
$$
f(m_i,\tau)=2m_i (2\pi i\tau/m_i)^{2/3} \exp (im_i\tau).  \eqno (2.4)
$$

In the formula (2.2) the expression $p_ix_i$ in the exponent is originally
divided by Planck's constant over $2\pi$. But that factor has been removed by
choosing units of space and time so that Planck's constant (divided by $2\pi$) 
and the velocity of light are both unity. But then letting $\tau$ go to 
infinity is effectively equivalent to letting Planck's constant go to zero: the
expansion of the spacetime scale is mathematically equivalent to going to the 
classical limit. Formula (2.3) shows that in this limit the probability 
distribution in spacetime for a freely moving particle is specified by the 
momentum-space distribution function $\psi_i(p_i)$ in accordance with the 
relativistic classical physics formula $p_i=m_iv$.

The fall-off property described above was derived from quantum theory.
Later I shall derive it from classical physics. 

The correspondence principle asserts that the classical-physics results
hold not only for these free-particle states but also for processes 
corresponding to networks of locally interacting particles that propagate 
freely over the asymptotically large distances between their interactions: 
the classical physics probabilities emerges from the quantum probabilities 
in the asymptotic $\tau \rightarrow \infty$ limit. This 
correspondence-principle requirement determines not only the locations 
and natures of the singulaties of the quantum momentum-space scattering 
functions, but normally entails also that, in a real neighborhood of a 
singular point $P$, the scattering function is a limit of a function 
analytic in the intersection of a complex neighborhood of $P$ with the 
interior of a cone that extends from the real domain in a set of directions 
that is specified by the structures of the classical scattering diagrams 
associated with that singular point $P$. This connection between 
{\it directions of analyticity} at singularities and classical spacetime 
diagrams is made via a $4n$-dimensional displacement vector $U$ introduced 
in reference [7].\\ 

{\bf 3. The $4n$-dimensional displacement vector $U$.}

Consider a spacetime diagram $D$ that describes a possible network of
classical particles with a total of $n$ initial and final particles. This 
diagram $D$ determines (via the directions of the initial and final lines)
a set $P=(p_1, ... ,p_n)$ of initial and final momentum-energy vectors. 

It is convenient to introduce in addition to the {\it physical} 
momentum-energy vectors $p_i$, which have positive energy components, also the
{\it mathematical} momentum-energy vectors $k_i$, where $k_i=p_i$ for initial
particles, and $k_i= -p_i$ for final particles. Then the law of conservation
of energy momentum reads $\sum k_i = 0$. 

The $4n$-dimensional displacement vector $U$ is defined as follows. 
From any arbitrarily chosen origin $O$ in spacetime draw, for 
each initial and final particle $i$, a vector $u_i$ from $O$ to some point 
on the straight-line that contains the initial or final line $i$. Define 
$$
U = (u_1, ... , u_n).    \eqno (3.1)
$$

For a fixed spacetime diagram D this 4n-dimensional displacement vector $U$ 
is not uniquely fixed: one can add to $U$ any vector of the form 
$$
U_0=(a+b_1k_1, a+b_2k_2, ... , a+b_nk_n), \eqno (3.2)
$$
where a is a real spacetime vector, and for each $i$ the parameter 
$b_i$ is a real number. Changing $a$ just shifts the location of $D$ 
relative to the origin $O$, and changing $b_i$ just slides the tip of 
$u_i$ along the straight line $i$.

Notice that the combination of the four conservation-law delta functions and
the $n$ mass-shell delta functions restricts the relevant set of points in 
the $4n$-dimensional space of points $K= (k_1, ... ,k_n)$ to a surface of 
co-dimension $4+n$, and that the $4+n$ dimensional set of vectors $U_0$ spans
the set of normals to that co-dimension $4+n$ surface: the contravarient 
vectors formed by taking linear combinations of the gradients to the arguments
of the $4+n$ delta functions constitute the set of vectors $U_0$. This is the
simplest example of the important fact that the set of vectors $U$ associated 
with a singular point $K$ generally span the space defined by the set of
normal vectors to the surface of singular points passing though $K$. 
This normality of the vectors $U$ associated with diagrams of classical 
physics to the surfaces of singularities of the S matrix provides the link 
between relativistic classical physics and domains of analyticty of 
scattering functions in relativistic quantum physics.

{\bf 4. Another Asymptotic Fall-Off Property.}

If the wave function $\psi_i(p_i)$ in Eq. (2.1) is infinitely differentiable
and of compact support, and if $V$ is the associated velocity (double) cone 
consisting of all lines through the origin ($p_i=0$) that intersect the 
compact support (in the mass shell $p_i^2=m_i^2$) of $\psi_i(p_i)$ then, for 
all $u$ in any compact set that does not intersect $V$, the function
$\widetilde\Psi (u\tau)$ uniformly approaches zero faster than any inverse 
power of the scale parameter $\tau$: for any integer $N$
$$ 
\lim_{\tau\rightarrow\infty}\tau^N \widetilde\Psi_i(u\tau)=0.     \eqno (4.1)
$$ 
This is a standard result (cf. ref[8], Eqn. (28)),
and it allows one to prove the weaker analyticity properties that hold 
modulo infinitely differential back-ground terms. (See ref. [7]). But
to derive full analyticity from the correspondence principle a stronger 
fall-off property is needed.

This stronger asymptotic fall-off property is obtained by introducing into
the wave functions $\psi_i(p_i)$ an exponential factor that
shrinks in width as $\tau$ tends to infinity. Specifically, one 
introduces free-particle momentum-space wave functions of the form 
$$
\psi_{\tau,\gamma,\bar p}(p) = \chi(p) \exp (-(p-\bar p)^2 \gamma\tau). 
\eqno (4.2)
$$  
and also requires the infinitely differential function $\chi(p)$ 
(of compact support) to be analytic at $p=\bar p$, where 
$p^2 = \bar p^2 = m^2.$  Then the following fall-off property holds:  
for all $4$-vectors $u$ in any compact set that does not intersect the line 
through the origin containing $\bar p$, and for all $\gamma \geq 0$ smaller 
than some fixed $\gamma_0$, there is a pair of finite numbers $(C,\alpha)$ 
such that for all $\tau$
$$
|\widetilde\Psi_{\tau,\gamma,\bar p}(u\tau)|< 
C\exp -\alpha\gamma\tau. \eqno (4.3)
$$ 

Classical and quantum proofs of this fall-off property will be described 
below. But let us first show how this property of the free-particle 
coordinate-space wave functions is used to deduce, from the correspondence 
principle, domains of analyticity for the momentum-space scattering function.

{\bf 5. Kinematics and Probabilities.}

The connection to the correspondence principle is obtained by using initial 
and final wave functions $\Psi_i(p_i,u_i)$ of the form 
$$  
\Psi_i(\tau,\gamma,\bar p_i;p_i,u_i)= 
\Psi_i(\tau,\gamma,\bar p_i;p_i)\exp iu_ip_i   \eqno (5.1) 
$$
where, for any $i$, in accordance with (2.1) and (4.2), 
$$
\Psi(\tau,\gamma,\bar p;p)=\psi_{\tau,\gamma,\bar p}(p)2\pi\delta (p^2-m^2)
$$
The wave function (5.1) represents the particle state 
obtained by translating the state represented by $\Psi_i$ by the spacetime
displacement $u_i$. The parameters $\gamma$ are taken to be the same 
for all $i$. It is convenient to use henceforth real $\chi_i(p_i)$, each of 
which is equal to one (unity) in some finite neighborhood of $\bar p_i$.

The correspondence-principle results are obtained by examining the 
$\tau \rightarrow \infty$ behaviour of the transition amplitude
$$
A(\tau)=S[\{\Psi_i(\tau,\gamma,\bar p_i; p_i, u_i\tau)\}] \eqno (5.2)
$$
where the right-hand side is
$$
\left[\prod_i \int (2\pi)^{-4} d^4k_i
\Psi_i(\tau,\gamma,\bar k_i;k_i)\right]S(K)\exp iKU\tau. 
$$

The absolute value squared of the complex number $A(\tau)$, times $f(\tau)$, 
is the transition probability associated with these states of the initial 
and final particles, and $f(\tau)$ is the inverse of the square of 
the product of the norms of the wave functions $\psi_i$ of (4.2). This factor
grows like $(\tau)^{3n}$, but this growth can be absorbed into a bound of the 
form $C exp -\alpha\gamma\tau$ by a slight adjustment of $C$ and $\alpha$.

{\bf 6. The Correspondence-Principle Condition.}

For any fixed $\bar K$ (with $\sum \bar k_i =0$ and, for each $i$, 
$\bar{k}_i^2 = m_i^2)$ there is a set $C(\bar K)$ of vectors $U$ such that each 
pair of $4n$-dimensional vectors $(\bar K, U)$ satisfies the Landau-Nakanishi 
conditions. This set $C(\bar K)$ includes the set $C_0(\bar K)$ consisting of 
all of the vectors $U_0$ of the form (3.2): each of these vectors $U_0$ 
specifies a classical-physics diagram D in which all of the initial and 
final particles pass through a single common point. Each of these vectors 
$U_0$ has a null (Lorentz) inner product with every tangent vector to 
--- i.e., with every infinitesimal displacement in --- the surface at 
$\bar K$ of singularities generated by the mass-shell and overall 
conservation-law delta functions. 

Suppose $C(\bar K) = C_0(\bar K)$. That would mean that, on the one hand, 
there are for the set  $\{\bar k_i\}$ of initial and final (mathematical) 
momentum-energy vectors specified by $\bar K$ no classical-physics diagrams 
except the trivial ones in which all the initial and final particles pass 
through a common point, and, on the other hand, according to the Feynman 
rules, no singularity of the quantum scattering function. But from the 
S-matrix point of view the Feynmam rules are suspect, because they come 
essentially from the physically meaningless continuous time evolution, 
and also lead to infinities. However, the general correspondence principle 
condition that the predictions of classical physics should emerge in the 
limit where Planck's constant goes to zero, or, equivalently, where $\tau$ 
goes to infinity, would seem to be an exceedingly plausible and secure 
condition. The analyticty of the scattering function at this point $\bar K$ 
is, in fact, a consequence of that correspondence condition. 
      
For any point $\bar K$ such that $C(\bar K) = C_0(\bar K)$
consider any $U$ that does not belong $C(\bar K)$. If $U$ does not belong 
to $C(\bar K)=C_0(\bar K)$ then for at least one of the $n$ particles
$i$ the component vector $U_i$ is not parallel to $\bar k_i$. But then 
the amplitude $A(\tau)$ will pick up an exponential fall-off factor of 
the kind shown in (4.3). These vectors $U$ cover a unit sphere in the 
$3n-4$-dimensional subspace that is normal to the $n+4$-dimentional
subspace $C(\bar K)$. Thus there will be a {\it least value} of $\alpha$ 
for the $U$'s on this (compact) unit sphere.

This uniform exponential fall-off over this unit sphere arises, 
in the classical computation, from the exponential fall off of the overlap 
of the probability functions of the initial and final particles: 
i.e., from the exponentially decreasing probability, as $\tau$ increases, 
for {\it all} of the initial and final particles to be in any single 
finite region of space-time that grows like the square root of $\tau$. 
In classical physics such an exponential decrease in this probability, 
coupled with the fact that the only classical scattering process that 
can carry the initial momentum-energies to the final momentum-energies is 
one where all the initial and final particle trajectories pass through 
some such growing space-time region entails a similar fall off of the 
transition probabilities: the probability for this kind of classical process 
to occur cannot grow faster than the product of the probabilities that 
the particle can all be in any such growing region. Thus the 
correspondence principle requires that transition amplitude 
$A(\tau)$ have the same sort of fall off as the one arising from the 
overlap of the wave functions. It will now be shown that this condition
entails the analyticity of the scattering function at this point 
$\bar K$ where $C(\bar K) = C_0(\bar K)$.

{\bf 7. Derivation of analyticity at trivial points.}

By a ``trivial point'' I mean a point $\bar K$ such that 
$C(\bar K) = C_0(\bar K)$: the only classical processes with
external momenta specified by $\bar K$ are the trivial single-vertex 
diagrams.

The set of Landau-Nakanishi surfaces that enter any bounded region of 
$K$ space has been shown to be finite [Ref. 10]. And each such surface is
confined to a co-dimension-one analytic manifold. Consequently, each trivial 
point $\bar K$ lies in an open neighborhood of such points.
 
Introduce a set of analytic coordinates $q$ in the $3n-4$-dimensional 
manifold in $K$-space restricted by the mass-shell and conservation-law
conditions near $\bar K$. Let the $q$ be a subset of the space components of 
the set of vectors $(k_i-\bar k_i)$, and let the $v$ associated with
any $q(K)$ in the neighborhood of $q(\bar K)=0$ be the corresponding 
$3n-4$ components of $U\tau$ mod $C_0(K)$, so that $KU\tau$ in (5.2) 
becomes $(-qv -\bar kv)$, where the metric $(1,1,1)$ is now used, and $v$ 
represents displacements away from the displacements that generate the trivial
single-vertex processes. Then the $A(\tau)$ in (5.2), times the (unimportant) 
phase factor $\exp(i\bar kv)$. can be written as
$$ 
T(v,r)=\int dq F(q)\exp(-r\mu(q)) \exp(-iqv),  \eqno (7.1) 
$$
where
$$
\mu(q) = \sum_i (k_i(q)-k_i(0))^2,  \eqno (7.2)
$$
$r= \gamma \tau$, and $F(q)$ is the scattering function times a factor that is
real, infinitely differentiable of compact support, and analytic at $q=0$, 
which is the $q$-space image of $\bar K$. A fall-off property of the form 
(4.3) is required to hold for all $\tau$ and all $0\leq\gamma \leq \gamma_0$, 
with $r=\gamma \tau$, and all $v=\hat{v}\tau$ with $|\hat{v}|=1$ . What needs 
to be proved is that this fall-off condition, together with the analogous 
rapid (faster than any power of $\tau$) fall off at $\gamma = 0$, entails the 
analyticity of $F(q)$ at $q=0$.  

This rapid fall off of the bounded $T(v,0)=T(\hat{v}\tau,0)$ for all 
unit vectors $\hat{v}$ means that $F(q)$ is the well-defined and infinitely 
differentiable Fourier transform:
$$
F(q)=(2\pi)^l\int dv\exp(iqv) T(v,0),    \eqno (7.3)
$$
where $l=3n-4$. To show that $F(q)$ is analytic at $q=0$ re-write this 
equation in the form
$$
(2\pi)^l F(q)=\int dv\exp(iqv)\times
$$
$$
\left[T(v,\gamma_0|v|)exp(\gamma_0|v|\mu(q))
-\int_0^{\gamma_0|v|} dr \frac{\partial}{\partial r}
[T(v,r)\exp(r\mu(q))]\right]. \eqno (7.4)
$$

Consider first the first term in the big brackets. The correspondence
principle requires the factor $T(v,\gamma_0|v|)$ to be bounded by 
$C\exp(-\alpha\gamma_0|v|)$. The function $\mu(q)$ is zero at $q=0$,
and hence the associated exponential growth is dominated by the fall-off 
factor for $q$ in a sufficiently small neighborhood of $q=0$, Indeed, this 
bound keeps the integral well defined and analytic for all $q$ in a small 
complex neighbor of $q=0$. Thus the contribution $F_1(q)$ to $F(q)$ coming 
from the first term in the big brackets is analytic at $q=0$.

To prove that this property holds also for the other contribution, $F_2(q)$,
substitute (7.1) into the second term in the big brackets. The 
$\partial/\partial r$ can be moved under the integral over $dq$ because
$F(q)$ is infinitely differentiable of compact support. This gives for the
integrand
$$
\exp(iqv)\frac{\partial}{\partial r}[T(v,r)\exp(r\mu(q))]=
$$
$$
\int dq' F(q')\exp(i(q-q')v)\exp(r(\mu(q)-\mu(q')))[\mu(q)-\mu(q')]
\eqno (7.5)
$$
Hefer's theorem [8] allows one to write 
$$
\mu(q)-\mu(q')=\rho(q,q')\cdot(q-q'),   \eqno (7.6)
$$
where $\rho$ is a vector whose the components $\rho_j$ (j=1,... ,3n-4)
are analytic in a product of domains around $q=0$, and $q'=0$. Then (7.5) 
becomes
$$
\exp(iqv)\frac{\partial}{\partial r}[T(v,r)\exp(r\mu(q))]=
Div_v [\exp(iqv)\exp(r\mu(q))H(q,v,r)],   \eqno (7.7)
$$
where $H(q,v,r)$ is the vector
$$
H(q,v,r)= -i\int dq' F(q')\exp(-iq'v)exp(-r\mu(q'))\rho(q,q').
\eqno (7.8)
$$
We may thus write
$$
(2\pi)^l F_2(q)=-\lim_{R\rightarrow \infty}
\left[\int_{|v|<R} dv \int_0^{\gamma_0|v|} dr 
Div_v [\exp(iqv)\exp(r\mu(q)) H(q,v,r)]\right].
\eqno (7.9)
$$

For fixed $R$ we can change the order of integration and perform first
an integration over $v$ for $r/\gamma_0 < |v| < R$. Then Gauss' theorem
gives the volume integral of the divergence as the difference of two surface
integrals, one at $|v|=r/\gamma_0$, the other at $|v|=R$. The estimates given
in Appendix IV of ref. [9] show that the contribution at $R$ vanishes as 
$R\rightarrow \infty$. The contribution at $|v|= r/\gamma_0$ integrated on
$r$ from $0$ to $\infty$ generates an integration over all $v$ with $r$
replaced by $|v|\gamma_0$, and a Jacobian factor $J(q)$ that is analytic at
$q=0$. Thus we obtain  
$$
(2\pi)^l F_2(q) = \gamma_0\int dv \exp(iqv) 
\exp(\gamma_0 |v| \mu(q)) \widehat v\cdot H(q,v, \gamma_0 |v|),  
\eqno (7.10)
$$
where $\widehat v = v/|v|$. This function $F_2(q)$ is analytic at $q=0$ 
for the same reasons that $F_1(q)$ was. This completes the proof, apart
from the straightforward calculations given in Appendix IV of reference [9].

Note that that (7.1), with $r=\gamma_0 \tau$, and (7.4) together with (7,10),
gives a generalization of the Fourier transformation theorem that incorporates
Gaussian factors. It gives, from the mathematical point of view, a localized 
version of the familiar connection between analyticity and exponential fall off
of the Fourier transform. From the physics point of view it gives a connection 
between the analyticity of the scattering functions of relativistic quantum 
theory and the results of classical physics that emerge from quantum theory in 
the classical limit where Planck's constant goes to zero. 

The analyticity of the scattering functions except on the Landau-Nakanishi
surfaces has thus been derived, by ``reverse engineering'' the correspondence 
principle: quantum properties have been deduced from classical 
properties, the correspondence principle, and the basic connection between
classical and quantum physics, namely the Fourier-transform connection between
the momentum-energy and the space-time displacements of freely moving
particles.  

{\bf 8. Derivation of Cone of Analyticity at Most Singular Points.}

A more complex category of points $\bar K$ consists of points $\bar K$
such that all of the spacetime diagrams corresponding to this $\bar K$ 
are the same apart from shifts in location or scale, but which differ from 
the simple single-vertex case except in the limit where the diagram is 
shrunk to a point. For any such point $\bar K$ the set $C(\bar K)$ consists of 
$C_0(\bar K)$ plus a single ray, $U(\bar K)$: the displacements 
along $U(\bar K)$ generate the displacements of the external lines of the 
diagram away from positions where they all intersect at a single point. 
[The argument can be extended to cover all points $\bar K$ such that all 
of the Landau-Nakanishi surfaces that contain $\bar K$ coincide with a single 
co-dimension-one Landau-Nakanishi surface, and hence all specify the same 
unique ray $U(\bar K)$.]

It is important that $U(\bar K)$ is a ray, not a full line: 
a displacement in the opposite direction does not give the locations of 
the external lines of a classically allowed process. (The intermediate 
particles would have to move backward in time, and carry the incoming 
positive energy backward in time.) Thus a compact set of displacements 
$U$ not in $C(\bar K)$, but confined to a space essentially normal to the set 
$C_0(\bar K)$, cannot now cover an entire sphere: there must be a hole in 
this compact set through which the single ray $U(\bar K)$ can pass. 

To deal with this case one can introduce the same set of local coordinates
$(q,v)$ as before, with $\hat{v}=v/|v|$, and let $\hat{v}(\bar K)$ be the 
point on the unit sphere $|v|=1$ that is the image in $|v|=1$ of $U(\bar K)$. 
Let $A(\bar K)$ be a compact set in $v$ space that lies in the unit sphere, 
and covers this sphere $|v|=1$ except for points in a small open spherical 
ball about the point $\hat{v}(\bar K)$. Let the points in this ball that lie 
also on the sphere $|v|=1$ be called $H(\bar K)$ (for Hole), so that each point
on $|v|=1$ lies either in $A(\bar K)$ or in $H(\bar K)$, but not in both. 

Choose the functions $\chi(p_i)$ in (4.2) so that their supports are small 
enough so that the point $\hat{v}(K)$ corresponding to each point $K$ in the 
support of the product of the $\chi(p_i)$s lies in a closed subset of the open
set $H(\bar K)$. Then for all points $\hat{v}=v/|v|$ in $A(\bar K)$ the 
function $T(v,\gamma |v|)$ will, by virtue of the correspondence principle, 
fall off faster than any power of $|v|$ for $\gamma = 0$, and like (4.3) for
$0 < \gamma \leq \gamma_0$. The problem is then to show that
the function $F(q)$ in (7.1) is the boundary value, in some real neighborhood
of $q=0$, of a function analytic in the intersection of a complex neighborhood
of $q=0$ with an open cone $Q$ in $Im$ $q$.      

To prove this, separate the $v$-space domain of integration in (7.3)
into two disjoint parts, $V(H(\bar K))$ and $V(A(\bar K))$, where the latter
consists of all rays from $v=0$ that pass through the closed set $A(\bar K)$
of points in the sphere $|v|=1$, and $V(H(\bar K))$ is the rest of $v$ space.
   
This separation of the space of integration of the (bounded-by-virtue-of- 
unitarity) function $T(v,0)$ into two parts separates $F(q)$ into two terms:
$$
F(q) = F_H(q) + F_A(q).   \eqno (8.1)
$$   
The imaginary part of $q$ in $F_H(q)$ is restricted to the open cone $Q$
in which $Im$ $qv > 0$ for all $v$ in a closed cone $V$ that contains the 
closure of $V(H(\bar K))$ in its interior, apart from the origin $v=0$. 
For these $q$ the exponential factor $\exp iqv$ in (7.3) get from $Im$ $q$ 
a factor $\exp -\alpha |Im$ $ q||v|$, where $\alpha > const > 0$. This means,
because $T(v,0)$ is bounded, that the integral is absolutely converent, 
and hence that $F_H(q)$ is analytic near $q=0$ for $Im$ $q$ in $Q$.

Most of the real points $q$ very near to $q=0$ are ``trivial'' points, 
of the kind considered in the preceding section. At those trivial points 
$q'$, the function $F(q') = F_H(q') + F_A(q')$ is analytic.  These two terms 
are taken at these points $q'$ to be just the contributions to $F(q)$ 
specified in (7.4) and (7.10) restricted to the regions $V(H(\bar K))$ and 
$V(A(\bar K))$ respectively. Both of these contributions are analytic in 
the intersection of some neighborhood of $q$ with the cone $Q$. Thus one 
can stay in the domain of analyticity by moving $Im$ $q$ slightly into 
the cone $Q$ in order pass to the other side of the surface of 
singularities that passes through $q=0$.  

A more elaborate presentation of this argument, and of its generalizations 
to more complex cases, can be found in references [7] and [8], and also 
in Iagolnitzer's book [11].

{\bf 9. Correspondence-Principle Asymptotic Fall Off.}

I have described some of the analytic consequences of the fall-off properties
(2.3), (4.1), and (4.3). I turn now to a fuller discussion of the roots 
of these fall-off properties in the correspondence to classical properties.

The statistical predictions of quantum mechanics correspond, at least in a 
formal way, to the predictions of classical {\it statistical} mechanics. 
In the latter theory one describes a system of $n$ particles at any time $t$ 
in terms of a function $\rho (x,p,t)$, which specifies how the probability 
is distributed over the points $(x,p)$ of ``phase space,'' where $x$ 
specifies the $3n$ coordinate variables and $p$ specifies the $3n$ 
momentum-space variables. Free-particle evolution keeps $p$ fixed and 
shifts the location $x_i$ of a particle of (rest) mass $m_i$ during a 
time interval $t$ to the location $x_i + tp_i/m_i$. For large $t$ the second
term dominates, and the coordinate-space probability function goes over to 
the momentum-space probability function, properly scaled to account for the
diverging directions of the different momentum vectors. This classical 
kinematics entails that for free particles the classical distribution 
$\rho (x,p,t)$ at large times $t$ becomes a product over $i$ of functions
$$
\rho (u_it, p_i,t)= |\rho (u_im_i) \rho(p_i)/f(m_i,t)^2|, \eqno (9.1)    
$$
where
$$
\rho(p_i)=\int d^3x_i \rho(x_i,p_i, t'),  \eqno (9.2)
$$ 
is independent of $t'$, and $f(m_i,t)$ is the function defined in (2.4). 
Here I am, for simplicity, assuming that the momenta are small enough so 
that the non-relativistic formulas (where $t=\tau$ and $p_0=m$) are adequate.
(The fully covariant formulation gives the same results.) The factor 
$(m_i/t)^3$ coming from $f(m_i,t)^{-2}$ compensates for the linear spreading 
out of the probability distribution in coordinate space, and the $1/(2m_i)^2$ 
comes from the normalization in (2.1). This equality of the classically-derived
and quantum-mechanically derived limits constitutes, in this case, 
{\it part of} the correspondence-principle relationship between the 
asymptotic properties in classical and quantum theory: both theories give 
the same asymptotic form for the probability distribution in $(x,p)$, 
for the case $\gamma =0$. 

There is no conflict here with the uncertainty principle limitation on the 
idea of a distribution in both $x$ and $p$ simultaneously: the huge 
spreading out of the coordinate-space distribution eliminates any such 
conflict. 

But what is the rate of approach to this limit?

The probability distribution in coordinate space at $t=0$ for the function 
in (4.2), at $\gamma=0$, would be given by the (absolute value squared of the) 
Fourier transform of $\chi(p)$. This transform of the infinitely 
differentiable compactly supported $\chi(p)$ falls off faster than any power 
of $|x|$. This leads to the quantum mechanical prediction (4.1). Classically, 
this original $x$-space distribution is the constant (non-expanding) 
background to the $t$-dependent diverging trajectories. If this non-expanding
background falls off faster than any power of $x$ then its contribution at 
points $x=u\tau$ will fall off faster than any power of $\tau$. Hence the 
approach to the large-$t$ limit computed classically, by using the 
straight-line trajectories in space-time, also exhibits the faster than 
any power fall off specified in (4.1): the classical and quantum predictions 
agree about both the limit and the rate of approach to this limit.    

But what is the rate of fall off for the case $\gamma >0$?

To show that the fall off in this case conforms to (4.3) it is sufficient to
go to the frame where $\bar p$ is pure spacelike and the space part of
$u$ is nonzero. Then
$$
|\widetilde{\Psi}_{\tau,\gamma,\bar p}(u\tau)|=
|\int d^3q/(2\pi)^{-3}\chi(q)\exp(-\tau[q^2\gamma+i(qu-u_0(q_0-\bar p_0))])|,
\eqno (9.3)      
$$
where I again use the metric (1,1,1) for the $3$-vector products $qu$ 
and $q^2$, and $q_0-\bar p_0 =(q^2+m^2)^{1/2} -m$.

To get the quantum prediction, consider a distortion of the $q$-space 
contour that is parameterized by a scalar $\alpha$. For $q^2 > \alpha$ 
there is no distortion. For $Re$ $q^2 < \alpha$  the component of 
$Im$ $q$ that is directed along $u$ is 
shifted (keeping real the other two components of the $3$-vector $q$) so that

$$ 
Re[q^2\gamma+i(qu-u_0(q_0-\bar p_0))]=\alpha\gamma. \eqno (9.4)
$$
Distort the contour from $\alpha =0$ to a value such that all real $q$
in $q^2 \leq \alpha$ lie inside the open set where $\chi$ is one,
and such that $|Im$ $q|$ remains less than $m$. 
  
Then for all real points $q$ with $ q^2>\alpha$ one has an exponential 
fall-off factor $\exp -\alpha\gamma\tau$. For real $q$ such that 
$q^2 < \alpha$ the condition (9.4) gives a factor $\exp -\alpha\gamma\tau$. 
One can obtain a bound like this for every four vector $u$ on the unit 
(Euclidean) sphere, minus small open holes around the rays along the 
positive and negative time axis (along which $\bar p$ has been taken to lie). 
These holes can be defined by conditions on the three-vector part $\vec{u}$ 
of $u$: $|\vec{u}|< \epsilon$. The only singularity that could block this 
continuation is the singularity of $q_0$ at $q^2+m^2=0$, and this is 
prevented by our condition $|Im$ $q|<m$. 

A more detailed presentation is given in Appendix III of Ref. 9.

The classical analog is obtained by taking the classical 
coordinate-space probability function, imagined now to specify the 
distribution of the classical particles, to be the one obtained from
the Fourier transform. For large $\tau$ the contributions from $\chi - 1$
fall off exponentially. Ignoring that contribution, at very large
$\tau$, one has a coordinate-space function that is essentially a 
Gaussian, which has a width that grows like the square root of $\tau$.
Hence in the scaled-down coordinate $u=x/\tau$ the width of the Gaussian 
shrinks like $(\tau)^{-1/2}$, just as it does in momentum space. 
Thus the probability function in $(u,p)$-space 
(or in $(\vec{u},\vec{p})$-space) for fixed $(u,p)$, falls off 
exponentially in $\tau$, as long as one keeps $|\vec{u}|$ finitely 
away from zero.

The fall-off properties (4.1) and (4.3) pertain to the individual freely
moving particles. But we need analogous fall-off properties for process 
involving multiple scatterings of such freely moving particles by 
quasi-local interactions. 
     
In quantum theory one has an initial $\Psi_{in}$ and a final $\Psi_{fin}$. 
If a certain preparation procedure $In$ prepares a system to be in the 
initial state $\Psi_{in}$, and if a certain measurement procedure $Fin$ 
will definitely produce a ``Yes'' outcome if the final state is $\Psi_{fin}$, 
and will definitely produce a ``No'' outcome if the final state is orthogonal 
to $\Psi_{fin}$, then 
$$ 
Probability = |\Psi_{in}^* S \Psi_{fin}|^2 \eqno (9.5) 
$$ 
is the predicted probability that a preparation of type $In$ followed by 
a measurement of type $Fin$ will yield an outcome ``Yes''. 

If the intersection of the supports of the wave functions (4.2) contain no 
points  $K$  such that $C(K)$ is bigger than $C_0(K)$ then the only
relevant classical scattering diagrams are the trivial one that have only
one vertex. If the interactions not carried by physical particles have finite
range (with perhaps exponential tails) then the transition probability
will (as mentioned previously) be bounded, in classical physics, by the 
probability that all of the particles can be in some region that grows like
the square root of $\tau$. And the condition that $C(K)= C_0(K)$ for all 
points in the support of the wave functions means that for any such growing
region in spacetime the probability that all the particles will be 
in this region will have an exponential in $\tau$ fall off coming from some
nonzero displacement in either a momentum variable $q$ or a translation 
variable $u$. And the range of these displacements is compact: they cover 
the compact surface in $v$ space times the product of the compact domains in
$q$-space. Thus for these ``trivial'' points one gets, in 
the classical-physics analog, a fall off of type (4.3), as already noted.

But how does one get the analogous result for multiple-scattering
processes, which involve intermediate particles? 

The answer is that if all interaction regions can be taken to grow no
faster than the square root of $\tau$, then in the scaled-down (by a factor 
$\tau$) coordinate system the diagram must have point vertices. And 
momentum-energy is strictly conserved in classical mechanics. So the 
scaled-down diagrams depict classical processes with point vertices. If no 
such diagram can match the external conditions imposed by the $(U,K)$ then 
there will always by an exponential fall-off factor coming from some external 
particle, which is what the arguments require.

{\bf 10. Nature of the Singularity.}

The correspondence principle entails analyticity except on the surfaces
specified by the Landau-Nakanishi equations, and it assures analyticity in
the associated cones of analyticity at the Landau-Nakanishi points. But 
what about the nature of these singularities?
 
Consider a $3$-particle to $3$-particle process in which two particles
collide to create one final particle plus one intermediate particle that
eventually collides with the third initial particle to produce the other 
two final particles. Classical physics demands that in the positive-time
asymptotic regime the transition probability function must fall off as 
$\tau^{-3}$, due to the geometric spreading. This is just the fall off 
obtained in section 2, and it corresponds to a pole singularity, 
$$
f(p)= i(p^2 - m^2+i\epsilon)^{-1},    \eqno (10.1)
$$
which is the energy-increases-with-time part of the mass-shell delta function
$2\pi\delta(p^2-m^2)$ of classical physics. Thus not only the location of 
this singulatity, and the $i\epsilon$ rule for continuing around it, but 
also the pole character of this singularity is determined essentially by 
the fall-off properties entailed by the correspondence principle. 

The geometric conditions that lead to the $\tau^{-3/2}$ fall off
in the single-intermediate-particle case can be generalized to the
case of any number of intermediate particles. One obtains the condition
$$
2d=3N_l -4(N_v -1)-1,   \eqno (10.2a)
$$
or
$$
d= \frac{1}{2}(3N_l -4N_v +3), \eqno (10.2b),
$$
where $N_v$ is the number of vertices, $N_l$ is the number of internal
lines, and $d$ is the ``degree'' of the singularity, with $d = -1$ 
being $\delta (E)$ or $E^{-1}$, and $d=0$ being $\log E$, etc. Thus for the 
two-vertex, one internal line case one gets $d=-1$ (a pole singularity)
and for the triangle diagram with three vertices and three internal lines 
one gets $d=0$ (a logarithmic singularity.) For $N_v=2$ and $N_l=2$ 
(two-particle threshold) one gets $d=1/2$, $(\sqrt E)$.

To understand (10.2) from the classical point of view consider the application
of (9.1), applied to the entire classical diagram $D$, consisting of $N_l$ 
internal lines, $N_e$ external lines, and $N_v$ vertices. The factors 
$|\rho(u_im_i)/f(m_i,t)^2|$, with $tp_i/m_i =\tau u_i$, give the $3N_l$ in 
(10.2a). Each internal lines contributes a factor $\tau^{-3}$ to the fall-off
of the probability, and hence a fall-off factor $\tau^{-3/2}$ in the amplitude,
and this translates via the Fourier connection to an increase by $3N_l/2$ of 
the degree $d$ of the singularity.

But the classical formula (9.1) has also a momentum factor $\rho(p)$.
The $p_i$ in (9.1) must include an external momentum-energy four-vector at 
each external line, and the function $\rho(p)$, with $p$ being the collection 
of internal and external four vectors, will have a conservation-law delta 
function at each of the $N_v$ vertices. This is a classical condition.
The scattering function has the one overall conservation-law
delta function factored off, leaving $4(N_v-1)$ delta functions. 

The term of zeroth order in $N_l$ and $N_v$ is not determined by this argument,
but is fixed by the known pole case to be the extra term $-1$ in (10.2a).
The important point is that to the extent that (10.2) determines the 
degree $d$ of the singularity, this degree is fixed by the fall-off
and conservation-law features exhibited by the associated classical process:
the classical process exhibits the features that enter into Eqn. (10.2).

These remarks tie Eqn. (10.2) to classical physics, but do not give a 
derivation of (10.2). This equation is derived in Kawai and Stapp [12], 
for all of the cases mentioned above, and, more generally, for each 
physical-region singularity that corresponds to a unique Landau-Nakanishi 
diagram in which no two vertices coincide, at most two lines connect any 
pair vertices, and no vertex is trivial in the sense that all of the lines 
connected to it are parallel. [Actually, far more is derived in ref. 12, 
namely an explicit form of the S-matrix near certain points where 
several surfaces intersect, and these forms play an important role in 
understanding the global analytic structure of the S matrix.] The proof is 
based on the analyticity properties derived from the correspondences principle,
on the general theory of holonomic microfunctions described in Sato, Kawai, 
and Kashiwara [13], and on the techniques and results developed in Coster 
and Stapp [14, 15] for combining the analytity properties that follow from 
the correspondence principles with the important unitarity property of the 
S matrix.

The other key element in S-matrix theory is ``crossing'': the
postulate that a certain analytic continuation that changes $k_i$
to $-k_i$ will take one to the scattering function of a ``crossed'' process
where initial (resp. final) particle $i$ is replaced by final (resp. initial)
anti-particle $i$. Hence much of the structure of quantum theory is seen to 
be entailed already by the correspondence principle, plus natural extensions 
of the analyticity properties entailed by the correspondence principle. 

{\bf 11. Photons and Infra-red Divergences.}

Massless particles, such as photons, pose new technical problems, which are 
entwined with an important infra-red problem. A number of studies [16, 17, 18]
of the effects of the interaction of an electron (or positron) with 
low-energy photons appeared to show that the pole-character of the electron 
is disrupted by this interaction: the pole exponent $-1$ is
modified by a term of order $1/137$. However, any such change at the level
of the S matrix itself would entail a significant deviation from the $1/r^3$ 
fall off, which is empirically confirmed to very high accuracy. 

Part of the problem in those works is that what was studied was the electron
{\it propagator}, which corresponds, physically, to suddenly creating
a charged electron at some point $x$ and suddenly destroying it at
some other point $x'$. But charge is conserved: it cannot be suddenly
created or destroyed. So one should examine, instead, closed loops of charge, 
where two particles of opposite charge emerge from an initial place, and
eventually come together at some later place. But even when this is done there 
still remains an infra-red divergence problem, associated with the emission
of ``infinite'' numbers of soft (i.e., low-energy) photons at each place where
some deflection or deviation of the spacetime trajectory of the charged 
particle from straight-line motion occurs. This infra-red problem is solved
by again appealing to the correspondence principle.

The point is this. If one considers the space-time diagram associated with 
the Feynman graph as a classical multiple-scattering process---of charged
particles---then one can compute the classical electromagnetic field
radiated by those moving charges. It has long been known that for every
classical radiation field there is a corresponding quantum state, called
a coherent state. It involves infinite numbers of photons. To resolve
the infra-red divergence problem completely one should use for the final 
quantum state of the radiant elecromagnetic field, not the vacuum state plus 
added photons, but rather the quantum coherent state corresponding to the 
classical electromagnetic field radiated from the classical process 
specified by the Landau-Nakanishi diagram, plus added photons. So again, 
as before, the quantum process is largely determined by the underlying 
classical process: the classical process determines the bulk of the 
radiated quantum electromagnetic field, and once this part is properly 
incorporated the fall-off properties associated with motions of the charged 
particles come into proper accord with the predictions of classical physics, 
which then fixes, via analyticty, the parts of the quantum scattering 
function closely associated with this classical process. One can then, again,
reverse engineer the correspondence principle to get the quantum counterpart 
of the classical process. The program was initiated by Stapp[19], and 
various resulting analyticity properties were derived in a series
of papers by Kawai and Stapp[20, 21]   

In the works described above the particle trajectories were always taken
to be straight-line segments. However, Eqn. (2.16) of ref. 19 shows the
effect of the ``Coulomb'' contribution. It conforms to the classical
rule. The correspondence principle approach discussed here suggests allowing 
the classical-particle trajectory to deviate from straight lines in a 
way that gives stationary action. That will cause these classical 
trajectories to curve as they do classically under the influence of a 
Coulomb potential. These curved trajectories will radiate soft photons that 
will need to be added to the final coherent state.

This suggested application of the correspondence principle begins to look 
more like a traditional spacetime description than an S-matrix calculation. 
However, it is built not upon the presumption of local interactions but rather
upon analyticity properties derived by a reverse engineering of the 
correspondence-principle classical limit.\\

{\bf References.}

\noindent 1. J. A. Wheeler, On the mathematical description of light nuclei
   by the method of resonating group structure. {\it Phys. Rev.}, {\bf 52}
   (1937), 1107-1122.\\
2. W. Heisenberg, Die ``beobachtbaren grossen'' in der theorie der 
   elementarteilen I and II. {\it Z. Physik}, {\bf 120} (1943), 513-538 
   and 673-702.\\
3. W. Heisenberg, ref. 2; G.F. Chew, {\it S-Matrix Theory of Strong 
   Interactions}.  W.A. Benjamin, New York, (1961); G.F.Chew, 
   {\it The Analytic S-Matrix}, W.A. Benjamin, New York, (1966). \\
4. L. D. Landau, On analytic properties of vertex parts. {\it Nucl. Phys.}, 
   {\bf 13} (1959), 181-192.\\
5. N. Nakanishi, Ordinary and anomalous threshholds in perturbation theory.
   {\it Prog. Theor. Phys.}, {\bf 22} (1959), 128-144.\\
6. S. Coleman and R. Norton, Singularities in the physical region. 
   {\it Nuovo Cimento}, {\bf 38} (1965), 438-442. \\
7. C. Chandler and H. Stapp, Macroscopic causlity conditions and 
   properties of scattering functions. {\it J. Math. Phys.}, {\bf 10} (1969), 
   826-859.\\
8. D. Iagolnitzer and H. Stapp, Macroscopic causality and physical region
   analyticity in S-matrix theory. {\it Commun. Math. Phys.}, {\bf 14} (1969), 
   15-55.\\
9. M Sato, Hyperfunctions and partial differential equations. {\it Proc. 
    Internat. Conf. on Functional Analysis and Related Topics, 1969},
    pp. 91-94, Univ. Tokyo Press. Tokyo, 1970.\\ 
10. H. Stapp, Finiteness of the number of positive-alpha Landau singularity
   surfaces in bounded portions of the physical region.
   {\it J. Math. Phys.}, {\bf 8}. (1967), 1606-1610.\\
11. D. Iagonitzer, {\it The S Matrix}. North-Holland, New York (1976).\\
12. T. Kawai and H. Stapp, Discontinuity formula and Sato's Conjecture, 
    {\it Publ. RIMS} {\bf 12} Suppl. (1977), 155-232.\\
13. M. Sato, T. Kawai and M. Kashiwara, Microfunctions and pseudo-
    differential equations. {\it Lecture Notes in Math.} No. 287, 
    pp. 265-529, Berlin-Heidelberg-New York, Springer, 1973.\\
14. J. Coster and H. Stapp, Physical region discontinuity equations. 
    {\it J. Math. Phys.}, {\bf 11} (1970), 2743-2763.\\
15. J. Coster and H. Stapp, Physical region discontinuity equations for
    multi-particle scattering amplitudes I. 
    {\it J. Math. Phys.}, {\bf 10} (1969), 371-396.\\
16. T. Kibble, Coherent soft-photon states and infrared divergences IV. 
    {\it Phys. Rev.}, {\bf 175} (1968), 1624-1640.\\
17. D.Zwanziger, Reduction formulas for charged particles and coherent states 
    in quantum electrodynamics. {\it Phys. Rev.}, {\bf D7} (1973), 1062-1098.\\
18. J.K. Storrow, Photons in S-matrix theory. {\it Nuovo Cimento}, 
    {\bf 54A} (1968), 15-41, and {\bf 57A} (1968), 763-776.\\
19. H.P. Stapp, Exact solution of the infrared problem. 
    {\it Phys. Rev.}, {\bf D28} (1983), 1386-1418.\\
20. T. Kawai and H. Stapp, Quantum electrodynamics at large distances. 
    {\it Phys. Rev.}, {\bf D52} (1995), 2484-2532.\\
21. T. Kawai and H. Stapp, On infra-red singularities associated with
    QC photons. {\it Microlocal Analysis and Complex Fourier Analysis}, World 
    Scientific Publishing, Singapore, 2002, pp. 115-134. 

\end{document}